\documentclass[twocolumn,showpacs,superscriptaddress,prl,aps]{revtex4}
\usepackage{amssymb}
\usepackage{amsmath}
\usepackage{amsfonts}
\usepackage{graphics}
\usepackage{epic}
\usepackage{eepic}
\usepackage{color}
\usepackage{epsfig}
\usepackage{latexsym}

\begin{document}

\def\ra{\rangle}
\def\la{\langle}
\def\be{\begin{equation}}
\def\ee{\end{equation}}
\def\nn{\nonumber}
\def\PRL{Phys. Rev. Lett.}
\def\PRA{Phys. Rev. A}
\def\NJP{New J. Phys.}

\def\X2{$\chi^{(2)}$}
\def\sinc{\textrm{sinc}}

\newcommand{\bra}[1]{\left\langle #1 \right\vert}
\newcommand{\ket}[1]{\left\vert #1 \right\rangle}

\title{Spectral Effects of Strong Chi-2 Non-Linearity for Quantum Processing}

\author{Patrick M. Leung}
\affiliation{Centre for Quantum Computer Technology, Department of
Physics,  University of Queensland, Brisbane 4072, Australia}

\author{William J. Munro}
\affiliation{Hewlett-Packard Laboratories, Filton Road, Stoke Giord,
Bristol BS34 8QZ, United Kingdom}
\affiliation{National Institute of Informatics, 2-1-2 Hitotsubashi,
Chiyoda-ku, Tokyo 101-8430, Japan}

\author{Kae Nemoto}
\affiliation{National Institute of Informatics, 2-1-2 Hitotsubashi,
Chiyoda-ku, Tokyo 101-8430, Japan}

\author{Timothy C. Ralph}
\affiliation{Centre for Quantum Computer Technology, Department of
Physics,  University of Queensland, Brisbane 4072, Australia}

\email{pmleung@physics.uq.edu.au}

\date{\today}

\pacs{03.67.Lx, 42.50.-p}

\begin{abstract}

Optical \X2 non-linearity can be used for parametric amplification and producing down-converted entangled photon pairs that have broad applications. It is known that weak non-linear media exhibit dispersion and produce a frequency response. It is therefore of interest to know how spectral effects of a strong \X2 crystal affect the performance. Here we model the spectral effects of the dispersion of a strong \X2 crystal and illustrate how this affects its ability to perform Bell measurements and influence the performance of a quantum gates that employ such a Bell measurement. We show that a Dyson series expansion of the unitary operator is necessary in general, leading to unwanted spectral entanglement. We identify a limiting situation employing periodic poling, in which a Taylor series expansion is a good approximation and this entanglement can be removed.

\end{abstract}

\maketitle

\section{I. Introduction}

An optical \X2 non-linearity can combine two lower energy photons into one higher energy photon via the parametric up-conversion process, and conversely, break a higher energy photon into two lower energy photons via parametric down-conversion, such that the total energy of the photons is conserved before and after the conversion. Optical \X2 non-linearity is widely used for parametric down conversion experiments to produce polarization entangled photon pairs, as well as for parametric amplification. Entangled photons have a large variety of applications, such as demonstration of Bell's inequalities violation~\cite{Kwiat95}, quantum error encoding~\cite{Ralph05}, production of heralded single photon sources~\cite{URen05}, quantum teleportation~\cite{Bennett93}, quantum dense coding~\cite{Bennett92}, and entanglement swapping~\cite{Pan98}. In addition, in principle a sufficiently strong \X2 non-linearity could be used to perform deterministic Bell measurement~\cite{Kim01}, as we shall discuss later in this paper. Bell measurements can be used in applications such as, teleporting qubits~\cite{Bennett93}, transferring quantum information with quantum repeaters~\cite{Briegel98}, as well as performing quantum computation~\cite{Gottesman99}. 

The conversion efficiency of current \X2 non-linearity is far below unity and past research has mostly concentrated on the properties of weak \X2. However, the strength of \X2 has been improving, as demonstrated in high photon number experiments, and it is therefore of increasing interest to examine the properties of strong \X2 media. Furthermore, since \X2 media have intrinsic spectral response, it is important to understand how the spectral effects of a \X2 medium affects the rate of up and down conversion. Spectral effects of weak \X2 crystals have been examined~\cite{Grice97} and some numerical research has been done for \X2 non-linearity of arbitrary strength~\cite{Raymer91}. In this paper, we shall model the spectral properties of a strong \X2 crystal and explore limits where analytical solutions can be found. To do so, we shall examine how the dispersion of a strong $\chi^{(2)}$ non-linear crystal affects the profiles of spectrally Gaussian input photons and determine how the probability of up-conversion depends on the dispersion. Specifically, here we consider Type II conversion. Moreover, we put forth a Bell measurement scheme based on parametric up-conversion and further develop a quantum gate from it, and examine the success rate of the gate under the influence of the dispersion of the crystal. Note that although our discussion focuses on up-conversion, a strong \X2 non-linearity has both up and down-conversion happening at the same time and thus the conditions and results that we derive here would also similarly apply to down-conversion.

This paper is arranged in the following way. The next section discusses how we model the spectral properties of a $\chi^{(2)}$ non-linear crystal. We indicate the problem of non-commutivity of the interaction Hamiltonian at different times, which requires us to use the Dyson series~\cite{Dyson, Sakurai} to calculate higher order effects of the crystal on the evolution of the photon states, as opposed to using the simpler Taylor series. The Dyson series leads to spectrally mixed states and this is undesirable for most applications. On the other hand, as we will show, spectrally separable solutions exist for the Taylor series. In section III, we examine the case where we send in a pair of separable photons through slices of weak $\chi^{(2)}$ crystal that are well separated and derive the output state for the photons and the probability of up-conversion. Then in section IV, we examine the case where we have periodically poled birefringent $\chi^{(1)}$ spacers in the crystal, and likewise derive the output state and the up-conversion rate. A rather surprising result of this paper is that we predict the efficiency of a strong bulk \X2 crystal, where the unitary evolution is modeled using the Dyson series, will be lower than the efficiency of the aforementioned thin slices case and the periodical poling case of a strong \X2 medium, where the unitary evolution is modeled using the Taylor series. In section V, we describe how one may construct a quantum gate based on Bell measurement with strong $\chi^{(2)}$ non-linearity and find the probability of success of the gate. We conclude in section VI. 

\section{II. Modelling the Spectral Properties of a Chi-2 Non-Linear Medium}
The Hamiltonian of a non-linear \X2 medium is spectrally dependent and it is crucial to understand how the spectral response of the medium affects the conversion rate, as well as the spectral profile of the output photons. So in this section, we shall model the spectral properties of a \X2 medium. The process of $\chi^{(2)}$ non-linearity can be studied in the interaction picture and the unitary evolution of a state vector is given by:

\begin{equation}
\hat{U}(t,t_0)|\psi\rangle=\exp\left(\mathcal{T}\Big{\{}\frac{1}{i\hbar}\int^{t}_{t_0}\hat{H}(t)dt\Big{\}}\right)|\psi\rangle
\label{eq:U}
\end{equation}

\noindent where $\mathcal{T}$ is the time ordering operator. Taking into account the time ordering leads to the following Dyson series~\cite{Dyson, Sakurai} expression for the unitary:

\begin{align}
\hat{U}(t,t_0)=&1+\frac{1}{i\hbar}\int_{t_0}^{t}\hat{H}(t_{1})dt_{1}\nn\\
&+\left(\frac{1}{i\hbar}\right)^{2}\int_{t_0}^{t}dt_{2}\int_{t_0}^{t_{2}}dt_{1}\hat{H}(t_{2})\hat{H}(t_{1})+\cdots\nn\\
&+\left(\frac{1}{i\hbar}\right)^{n}\int_{t_0}^{t}dt_{n}\int_{t_0}^{t_{n}}dt_{n-1}\cdots\int_{t_0}^{t_{3}}dt_{2}\int_{t_0}^{t_{2}}dt_{1}\nn\\
&\hat{H}(t_{n})\hat{H}(t_{n-1})\cdots\hat{H}(t_{2})\hat{H}(t_{1})+\cdots
\label{eq:Dyson}
\end{align}

If the interaction Hamiltonian commutes at different times, then the time ordering operator in equation~(\ref{eq:U}) has no effect and can be dropped, resulting in the usual Taylor series for the unitary expansion. The interaction Hamiltonian for a $\chi^{(2)}$ process has the form:

\begin{equation}
	\hat{H}(t)=\chi^{(2)}\epsilon_{0}\int_{V}d\mathbf{r}^{3}\hat{E_p}^{\dagger}(\mathbf{r},t)\hat{E_s}(\mathbf{r},t)\hat{E_i}(\mathbf{r},t)+h.c.
\end{equation}

Hereafter, we shall simplify the analysis to one spatial dimension, the propagation direction. This is legitimate if we consider collinear Type II conversion or in the case that the net transversal effects are negligible. The expression for the electric field operator of mode $j$ with spatial degree $z$ is:

\begin{equation}
\hat{E}^{\dagger}_{j}(z,t)=\int^{\infty}_{-\infty}d\omega_{j}A_j(\omega_j)\hat{a_j}^{\dagger}(\omega_j)\exp(i
k_j(\omega_j)z-\omega_j t)
\end{equation}

\noindent where $A_j(\omega_j)=i\sqrt{\frac{\hbar\omega_j}{4\pi c\epsilon_0
n^2_j(\omega_j)S}}$, $n_j(\omega_j)$ is the refractive index for mode $j$ and $S$ is the cross section area of the beam. We assume that $A_j(\omega_j)=A_j$ is slowly varying for the
frequencies of interest, allowing it to be factored outside the
integral. The frequency integrals have lower bounds extended from zero to negative infinity. This is mathematically legitimate because we are considering a system that operates at high frequency, where essentially there is no population present at low frequency. When we further integrate the Hamiltonian over $z$ from 0 to $L$, we obtain:

\begin{align}
	\hat{H}(t)=&\chi L\iiint_{-\infty}^{\infty} d\omega_{p}d\omega_{s}d\omega_{i}\hat{a}_{p}^{\dagger}(\omega_{p})\hat{a}_{s}(\omega_{s})\hat{a}_{i}(\omega_{i})\nn\\
	& \sinc\left(\frac{L\Delta k}{2}\right)e^{\frac{i L\Delta k}{2}}e^{-i\Delta\omega t}+h.c.
\label{eq:H1}
\end{align}

\noindent where $\Delta k=k_p(\omega_p)-k_s(\omega_s)-k_i(\omega_i)$ is the phase mismatch and $\Delta\omega=\omega_{p}-\omega_{s}-\omega_{i}$ is the frequency detuning. Here $\chi$ is again the interaction strength but incorporated with some constants from the electric field expressions. Following Grice and Walmsley~\cite{Grice97}, we Taylor expand the phase mismatch and retain terms up to first order by assuming higher order terms are negligible, then $\Delta k\approx \Delta k^{(0)} + k_{p}'\nu_p-k_{s}'\nu_{s}-k_{i}'\nu_{i}$, where $\nu_j=\omega_j-\mu_j$ and $\mu_j$ is the centre frequency of the photon in mode $j$. We set $\mu_s=\mu_i=\mu$ and $\mu_p=2\mu$. The parameter $k_{j}'$ is the derivative of wavenumber $k_j$ with respect to $\omega_j$ and evaluated at $\mu_j$. Due to conservation of momentum, the zeroth order term $\Delta k^{(0)}=k_{p}(\mu_p)-k_{s}(\mu_s)-k_{i}(\mu_i)=0$ and thus $\Delta k\approx k_{p}'\nu_p-k_{s}'\nu_{s}-k_{i}'\nu_{i}$. In our calculation, we assume that $k_{s}'-k_{i}'\ne0$, which is the case of Type II parametric conversion.

It can be shown that the interaction Hamiltonian in equation~(\ref{eq:H1}) does not commute at different times and therefore the Dyson series applies, instead of the Taylor series, when calculating higher order terms in the unitary expansion. The first order terms are identical, so we shall quantify and compare the similarity between the second term of the Taylor series and the Dyson series. A single photon with a Gaussian spectral profile is given by:

\begin{equation}
	|1\rangle=\int_{-\infty}^{\infty} d\omega_j~\hat{a}^{\dagger}(\omega_j)f(\omega_j)|0\rangle
\end{equation}

\noindent where $f(\omega_j)=\sqrt{\frac{1}{\sigma\sqrt{2\pi}}}\exp\left(-\frac{\nu_j^{2}}{4\sigma^{2}}\right)\exp\left(i\xi\nu_j\right)$ and $\la1|1\ra=\int|f(\omega_j)|^2d\omega_j=1$. Hence for up-conversion, the spectrally separable two-single-photon input state with modes $s$ and $i$ is: 

\begin{equation}
	|\psi_0\rangle=\iint_{-\infty}^{\infty} d\omega_{s}d\omega_{i}\hat{a}_{s}^{\dagger}(\omega_{s})\hat{a}_{i}^{\dagger}(\omega_{i})f(\omega_s)f(\omega_i) |0\rangle
\end{equation}

\noindent Let $|\psi_{2,T}\ra$ and $|\psi_{2,D}\ra$ be respectively the state components arising from the second term of the Taylor series and Dyson series expansions of the unitary operator acting upon the input state. Mathematically, 

\begin{align}
|\psi_{2,T}\ra =&\frac{1}{2!}\left(\frac{1}{i\hbar}\right)^{2}\left(\int_{-\infty}^{\infty}\hat{H}(t)dt\right)^{2}|\psi_0\ra\nn\\
=&\frac{1}{2!}\left(\frac{XL}{i\hbar}\right)^{2}A\iint_{-\infty}^{\infty}d\omega_{s}d\omega_{i}\hat{a}_{s}^{\dagger}(\omega{}_{s})\hat{a}_{i}^{\dagger}(\omega_{i})\nn\\
&\int_{-\infty}^{\infty}dt\left(at^{2}+bt+c\right)|0\rangle
\label{eq:psi2T}
\end{align}

\begin{align}
|\psi_{2,D}\ra=&\left(\frac{1}{i\hbar}\right)^{2}\int_{-\infty}^{\infty}dt_{2}\int_{-\infty}^{t_{2}}dt_{1}\hat{H}(t_{2})\hat{H}(t_{1})|\psi_0\ra\nn\\
=&\left(\frac{XL}{i\hbar}\right)^{2}A\iint_{-\infty}^{\infty}d\omega_{s}d\omega_{i}\hat{a}_{s}^{\dagger}(\omega{}_{s})\hat{a}_{i}^{\dagger}(\omega_{i})\nn\\
&\int_{-\infty}^{\infty}dt\left(at^{2}+bt+c\right)\left(\frac{1+Erf(dt+ig)}{2}\right)|0\rangle
\label{eq:psi2D}
\end{align}

\noindent where $A=\sqrt{\frac{\pi\gamma^{2}\sigma^{2}\left(k_{i}'+k_{s}'\right)^{2}}{2\left(k_{i}^{'2}+k_{s}^{'2}\right)}}$, \hspace{0.7cm}
$a=\frac{\sigma\left(k'_{i}-k'_{s}\right)^{2}}{k_{i}^{'2}+k_{s}^{'2}}$,\\ 

\hspace{0.5cm} $b=\frac{i\left(k'_{i}-k'_{s}\right)\left(k'_{i}\nu_{s}-k'_{s}\nu_{i}\right)}{k_{i}^{'2}+k_{s}^{'2}}$, \hspace{0.5cm} $c=\frac{\left(k'_{s}\nu_{s}+k'_{i}\nu_{i}\right)^{2}}{4\sigma^{2}\left(k_{i}^{'2}+k_{s}^{'2}\right)}$,\\ 

\hspace{0.5cm} $d=\frac{\sigma^{2}\left(k'_{i}-k'_{s}\right)^{2}}{\sqrt{2}\left(k_{i}^{'2}-k'_{i}k'_{s}+k_{s}^{'2}\right)}\sqrt{\frac{k_{i}^{'4}+k_{i}^{'3}k'_{s}+k'_{i}k_{s}^{'3}+k_{s}^{'4}}{\sigma^{2}\left(k'_{i}-k'_{s}\right)^{2}\left(k_{i}^{'2}+k_{s}^{'2}\right)}}$, \hspace{0.5cm}and\\ 

\hspace{0.5cm} $g=\frac{\left(k'_{i}-k'_{s}\right)^{2}\left(k'_{s}\nu_{s}+k'_{i}\nu_{i}\right)}{2\sqrt{2}\left(k_{i}^{'3}+k_{s}^{'3}\right)}\sqrt{\frac{k_{i}^{'4}+k_{i}^{'3}k'_{s}+k'_{i}k_{s}^{'3}+k_{s}^{'4}}{\sigma^{2}\left(k'_{i}-k'_{s}\right)^{2}\left(k_{i}^{'2}+k_{s}^{'2}\right)}}$. 

\noindent To quantify the similarity of the states $|\psi_{2,T}\ra$ and $|\psi_{2,D}\ra$, we define their fidelity to be:
\begin{equation}
F_{2}=\Big{|}\frac{\la\psi_{2,T}|\psi_{2,D}\ra}{\sqrt{|\la\psi_{2,T}|\psi_{2,T}\ra||\la\psi_{2,D}|\psi_{2,D}\ra|}}\Big{|}^2
\label{eq:F2}
\end{equation}

For the set of reasonable parameters, $k'_s=5.6\times10^{-9} (s/m)$, $k'_i=5.2\times10^{-9} (s/m)$, and $\sigma=10^9 (Hz)$, and assuming the extended phase matching condition $k_{p}'=\frac{k_{s}'+k_{i}'}{2}$, and the special condition $L^{2}\gamma\sigma^{2}\left(k_{s}'-k_{p}'\right)\left(k_{p}'-k_{i}'\right)=\frac{1}{2}$, we obtain $F_2=0.747$, 
which means the second term of the Dyson series significantly differs from the second term of the Taylor series. From equations~(\ref{eq:psi2T}) and~(\ref{eq:psi2D}), it is clear that this difference comes from the Erf function in $|\psi_{2,D}\ra$. By assuming the extended phase matching condition and the special condition, $|\psi_{2,T}\ra$ can become spectrally separable~\footnote{Not surprisingly, these two conditions are the same as the ones that Grice and Walmsley~\cite{Grice97} derived for spectrally separable parametric down conversion photons.} and proportional to $|\psi_0\ra$, however, the Erf function induces spectral entanglement between the two photons of $|\psi_{2,D}\ra$, which makes the two states substantially different. In principle, it is possible to test this difference experimentally by examining the efficiency difference between a moderately strong bulk \X2 crystal and for the same \X2 crystal being cut into many thin slices and separated sufficiently apart. We shall discuss the latter in the next section.

From the complexity of the 2nd order term of the Dyson series, we doubt that the high order terms in the series can have the spectral entanglement canceled out. Besides, for terms higher than the first order, the upper-bound of the time integrals is a time variable that has to be integrated by the next time integral. This makes the calculation of the sum of the Dyson series very difficult and complicated. If $\chi^{(2)}$ is sufficiently weak, then the higher order terms may be neglected and the calculation is tractable in this limit. For instance, Grice and Walmsley~\cite{Grice97} have examined weak parametric down conversion by ignoring higher order terms in the series. However, here we are interested the spectral effects of a strong $\chi^{(2)}$ crystal and we must include higher order terms for calculating the evolution of the input states. In the next section we shall look at the case where the strong $\chi^{(2)}$ crystal is cut into thin slices and separated sufficiently apart. Since the interaction is weak, we may legitimately ignore higher order terms for the evolution of the state provided by each thin slice.

\section{III. Obtaining Strong Chi-2 Non-Linearity from Many Weak Slices}

In this section, we examine the case where a strong \X2 non-linearity is obtained from many thin slices separated sufficiently apart, such that the wavepacket exits one slice before entering another and each slice provides only a weak interaction. In practice, this may be accomplished by having the photons in each mode passing one thin slice for many times in a loop. If the total length of a bulk \X2 crystal is $NL$ and we divide it into $N$ pieces of equal length $L$, then the unitary operator can be re-expressed as:

\begin{align}
\hat{U}(t_1,t_0)=&\lim_{N\rightarrow\infty}\left(1+\frac{1}{i\hbar}\int^{t_1}_{t_1-\Delta T}\hat{H}(t)dt\right)\nn\\
&\left(1+\frac{1}{i\hbar}\int^{t_1-\Delta T}_{t_1-2\Delta T}\hat{H}(t)dt\right)\dots\nn\\
&\left(1+\frac{1}{i\hbar}\int^{t_0+\Delta T}_{t_0}\hat{H}(t)dt\right)
\label{eq:U2}
\end{align}

\noindent where $\Delta T=(t_1-t_0)/N$. Each factor in the expression represents a weak interaction by a slice of crystal. Therefore, if the slices are sufficiently apart, the time bounds of the integrals can be conveniently extended to $-\infty$ and $\infty$ by considering far field limits~\cite{Giovannetti02}, and thus each factor in the expression is the same. The unitary operator can now be expressed as a Taylor series and the time ordering operator becomes irrelevant:

\begin{align}
\hat{U}=&1+\frac{N}{i\hbar}\int_{-\infty}^{\infty}\hat{H}(t)dt+\frac{1}{2!}\left(\frac{N}{i\hbar}\int_{-\infty}^{\infty}\hat{H}(t)dt\right)^{2}+\cdots\nn\\
=&\exp\left(\frac{N}{i\hbar}\int_{-\infty}^{\infty}\hat{H}(t)dt\right)
\label{eq:Taylor}
\end{align}

\noindent By integrating the interaction Hamiltonian in equation~(\ref{eq:H1}) over time $t$, we have:
\begin{align}
	\hat{H}=&\int^{\infty}_{-\infty}\hat{H}(t)dt\nn\\
	=&\chi L\iiint_{-\infty}^{\infty} d\omega_{p}d\omega_{s}d\omega_{i}\hat{a}_{p}^{\dagger}(\omega_{p})\hat{a}_{s}(\omega_{s})\hat{a}_{i}(\omega_{i})\nn\\
	& \sinc\left(\frac{L\Delta k}{2}\right)\exp\left(\frac{i L\Delta k}{2}\right)\delta(\Delta\omega)+h.c.\nn\\
	=&\hat{H}_{+}+\hat{H}_{-}
\label{eq:H2}
\end{align}

\noindent where we define 

\begin{align}
\hat{H}_{+}=&\chi L\iiint_{-\infty}^{\infty} d\omega_{p}d\omega_{s}d\omega_{i}\hat{a}_{p}^{\dagger}(\omega_{p})\hat{a}_{s}(\omega_{s})\hat{a}_{i}(\omega_{i})\nn\\
	& \Phi(\omega_{s},\omega_{i})\delta(\Delta\omega)
\end{align}

\begin{align}
\hat{H}_{-}=&\chi^{*} L\iiint_{-\infty}^{\infty} d\omega_{p}d\omega_{s}d\omega_{i}\hat{a}_{p}(\omega_{p})\hat{a}_{s}^{\dagger}(\omega_{s})\hat{a}_{i}^{\dagger}(\omega_{i})\nn\\
	& \Phi^{*}(\omega_{s},\omega_{i})\delta(\Delta\omega)
\end{align}

\begin{align}
\Phi(\omega_{s},\omega_{i})=& \sinc\left(\frac{L\Delta k}{2}\right)\exp\left(\frac{i L\Delta k}{2}\right)
\label{eq:Phi}
\end{align}

Using equations~(\ref{eq:Taylor}) to~(\ref{eq:Phi}), we shall now derive the evolution of the input state $|\psi_0\ra$. To calculate the first order term of the Taylor series, we act the Hamiltonian from equation~(\ref{eq:H2}) onto the input state, which gives:

\begin{equation}
\frac{N}{i\hbar}\hat{H}|\psi_{0}\rangle=\frac{N}{i\hbar}\hat{H}_{+}|\psi_{0}\rangle
=\frac{\chi NL}{i\hbar}\int d\omega_{p}\hat{a}_{p}^{\dagger}(\omega_{p})J_{p}|0\rangle
\end{equation}

where $J_{p}=\int d\omega_{s}f(\omega_{s})f(\omega_{p}-\omega_{s})\Phi(\omega_{s},\omega_{p}-\omega_{s})$. Similarly, the second order term of the Taylor series is:

\begin{align}
&\frac{1}{2!}\left(\frac{N}{i\hbar}\right)^{2}\hat{H}^{2}|\psi_{0}\rangle\nn\\
=&\frac{1}{2!}\left(\frac{N}{i\hbar}\right)^{2}\hat{H}_{-}\hat{H}_{+}|\psi_{0}\rangle\nn\\
=&\frac{1}{2!}\left(\frac{\chi NL}{i\hbar}\right)^{2}\iint d\omega_{s}d\omega_{i}\hat{a}_{s}^{\dagger}(\omega_{s})\hat{a}_{i}^{\dagger}(\omega_{i})\Phi^{*}(\omega_{s},\omega_{i})J_{s,i}|0\rangle
\end{align}

where we further define:
\begin{equation}
J_{s,i}=\int d\omega f(\omega)f(\omega_{s}+\omega_{i}-\omega)\Phi(\omega,\omega_{s}+\omega_{i}-\omega)\nn
\end{equation}

Likewise the third order term of the Taylor series is:

\begin{align}
&\frac{1}{3!}\left(\frac{N}{i\hbar}\right)^{3}\hat{H}^{3}|\psi_{0}\rangle\nn\\
=&\frac{1}{3!}\left(\frac{N}{i\hbar}\right)^{3}\hat{H}_{+}\hat{H}_{-}\hat{H}_{+}|\psi_{0}\rangle\nn\\
=&\frac{1}{3!}\left(\frac{\chi NL}{i\hbar}\right)^{3}\int d\omega_{p}\hat{a}_{p}^{\dagger}(\omega_{p})R_{p}J_{p}|0\rangle\hspace{2cm}
\end{align}

\noindent where $R_{p}=\int d\omega|\Phi(\omega,\omega_{p}-\omega)|^{2}$. Each of these terms contains the function $J$, which is obtained by integrating the product of the Gaussian spectral profiles of the photons and the sinc frequency response function of the \X2 medium.  Since it is difficult to integrate products of Gaussians and sinc functions, in our calculation, we have made use of the approximation, $\sinc(x)\approx\sqrt{\gamma\pi}\exp(-\gamma x^2)$, where the parameter $\gamma\approx0.193\dots$, is derived from equating the full-width-half-maximum of the two functions. After integrating the expression of $R_p$, we get $R_{p}=\sqrt{\frac{2\gamma\pi^3}{L^{2}\left(k_{s}'-k_{i}'\right)^{2}}}$, which is independent of $\omega_{p}$. Thus we can drop the $p$ subscript and set $R=R_p$. 

Lets suppose $|\psi_{even}\ra=\cos(N\hat{H}/\hbar)|\psi_0\ra$ and $|\psi_{odd}\ra=\sin(N\hat{H}/\hbar)|\psi_0\ra$ are, respectively, the sum of the even and odd terms of the unitary expansion after acting on the input state, such that the output state is $|\psi_{out}\ra=\hat{U}|\psi_0\ra=|\psi_{even}\ra -i |\psi_{odd}\ra$. The odd state represents the part of the output state in which the two photons in the signal and idler modes are up-converted into a photon in the pump mode. The even state represents the part of output state in which the two photons in the signal and idler modes are not up-converted and remain in the two modes. From the definition of the odd state, we obtain:
 
\begin{align}
|\psi_{odd}\ra=&\sin\left(\frac{N}{\hbar}\hat{H}\right)|\psi_0\ra\nn\\
=& e^{i\theta}\int d\omega_{p}\hat{a}_{p}^{\dagger}(\omega_{p})\frac{J_{p}}{\sqrt{R}}\sin\left(\frac{|\chi|NL}{\hbar}\sqrt{R}\right)|0\rangle\nn\\
=& e^{i\theta}\frac{B}{\sqrt{R}}\sin\left(\frac{|\chi|NL}{\hbar}\sqrt{R}\right)|\psi_p\ra
\label{eq:psiOdd}
\end{align}

\noindent where $B=\sqrt{\frac{\left(2\pi\right)^{3/2}\gamma\sigma_{p}}{2+L^{2}\gamma\sigma^{2}\left(k_{s}'-k_{i}'\right)^{2}}}$, $|\psi_p\ra=\int d\omega_{p}\hat{a}_{p}^{\dagger}(\omega_{p})f_p|0\rangle$, $f_p=\sqrt{\frac{1}{\sqrt{2\pi}\sigma_p}}\exp(-\frac{\nu_p^2}{4\sigma_p^2})$, $\theta$ is the argument of the complex number $\chi$ in polar form,  and $\sigma_{p}=\sqrt{\frac{\sigma^{2}\left(2+L^{2}\gamma\sigma^{2}\left(k_{s}'-k_{i}'\right)^{2}\right)}{1+L^{2}\gamma\sigma^{2}\left(\left(k_{s}'-k_{p}'\right)^{2}+\left(k_{i}'-k_{p}'\right)^{2}\right)}}$.   Similarly, the definition of the even state gives:

\begin{align}
|\psi_{even}\ra=&\cos\left(\frac{N}{\hbar}\hat{H}\right)|\psi_0\ra\nn\\
=&\Big{(}1-\frac{1}{2!}\left(\frac{|\chi|NL}{\hbar}\right)^2\iint d\omega_{s}d\omega_{i}\hat{a}_{s}^{\dagger}(\omega_{s})\hat{a}_{i}^{\dagger}(\omega_{i})\nn\\
&\Phi^{*}(\omega_{s},\omega_{i})J_{s,i}+\cdots\Big{)}|0\rangle
\end{align}

Note that if $k_{i}'\rightarrow k_{s}'$, then $R=\sqrt{\frac{2\gamma\pi3}{L^{2}\left(k_{s}'-k_{i}'\right)^{2}}}\rightarrow\infty$, and the sin function in the odd state will have an unphysical infinite oscillation. This is caused by ignoring higher order terms of $\Delta k$. When $k_{i}'\rightarrow k_{s}'$, the contribution from higher order terms of $\Delta k$ becomes more significant and the infinity will be prevented if these terms are included. Nevertheless, in the following, we shall show that the probability goes to zero when $k_{i}'\rightarrow k_{s}'$, which means Type II conversion simply does not work in that limit and thus there is no observable physical problem. 

With the expression for $|\psi_{odd}\ra$, we can now determine the probability of having a pump photon by calculating $P(odd)=\Big{|}\frac{\la\psi_{out}|\psi_{odd}\ra}{\sqrt{\la\psi_{odd}|\psi_{odd}\ra}}\Big{|}^2=|\la\psi_{odd}|\psi_{odd}\ra|$, which gives:

\begin{equation}
P(odd)=\frac{B^2}{R}\sin^2\left(\frac{|\chi|NL}{\hbar}\sqrt{R}\right)
\end{equation}

\begin{figure}[h]
\centerline{\psfig{figure=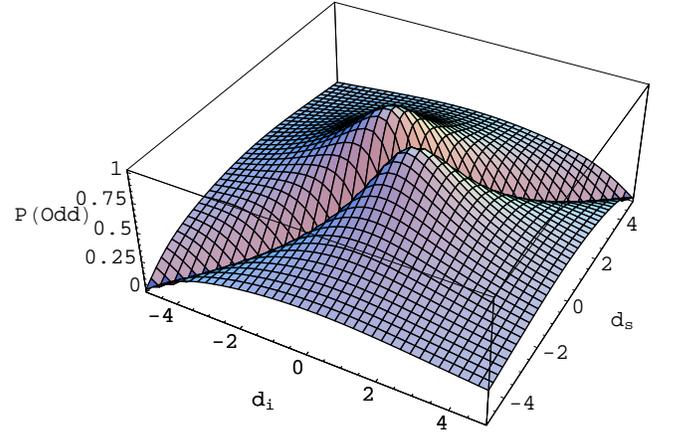,width=8.5cm}}
\caption{(Color online) Plot of the success rate in converting two photons, one in the signal mode and one in the idler mode, into a single photon in the pump mode using strong Type II parametric up-conversion, where $\frac{|\chi|NL}{\hbar}\sqrt{R}=\frac{\pi}{2}$. The rate is plotted against dimensionless parameters $d_{i}$ and $d_{s}$.\\} \label{fig:Podd}
\end{figure}

\noindent Setting $\frac{|\chi|NL}{\hbar}\sqrt{R}=\frac{\pi}{2}$ gives:

\begin{equation}
P(odd)=\frac{B^2}{R}=\sqrt{\frac{4\left(d_{s}-d_{i}\right)^{2}}{\left(1+d_{s}^{2}+d_{i}^{2}\right)\left(2+\left(d_{s}-d_{i}\right)^{2}\right)}}
\end{equation}

\noindent where $d_{s}=L\sqrt{\gamma}\sigma\left(k_{s}'-k_{p}'\right)$ and $d_{i}=L\sqrt{\gamma}\sigma\left(k_{i}'-k_{p}'\right)$. Figure~\ref{fig:Podd} shows the plot of $P(odd)$ against $d_{s}$ and $d_{i}$. The maximum probability is one and occurs at two points, $d_{i}=-d_{s}=\pm\frac{1}{\sqrt{2}}$, which is achieved when we have the extended phase matching condition $k_{p}'=\frac{k_{s}'+k_{i}'}{2}$, as well as the special condition $L^{2}\gamma\sigma^{2}\left(k_{s}'-k_{p}'\right)\left(k_{p}'-k_{i}'\right)=\frac{1}{2}$. The plot also shows that the probability is zero at $d_{i}=d_{s}$, that is when $k_{s}'=k_{i}'$. Beware that the zero probability trough is only meaningful in the sense that $k_{s}'$ is close to $k_{i}'$ but not equal. This is because, strictly speaking, in order to understand what happens at $k_{s}'=k_{i}'$, we have to include higher order terms of $\Delta k$ when calculating the probability. 

At the optimal points, the extended phase matching condition and the special condition are satisfied, and as we mentioned in the previous section, these conditions allow the second term of the Taylor series to be spectrally separable and proportional to $|\psi_0\ra$. In fact, these two conditions make $\Phi^{*}(\omega_s,\omega_i)J_{s,i}=Rf(\omega_s)f(\omega_i)$, and thus all even order terms in the series are spectrally separable and proportional to $|\psi_0\ra$, such that $|\psi_{even}\ra=\cos\left(\frac{|\chi|NL}{\hbar}\sqrt{R}\right)|\psi_0\ra$. From equation~(\ref{eq:psiOdd}), we can see that all the odd terms are proportional to $|\psi_p\ra$ as $|\psi_{odd}\ra=e^{i\theta}\sin\left(\frac{|\chi|NL}{\hbar}\sqrt{R}\right)|\psi_p\ra$. Physically this means that the evolution is a Rabi oscillation between the only two possible basis states $|\psi_p\ra$ and $|\psi_0\ra$. So inside the crystal, two processes happen concurrently, the two photons are up-converted into a pump photon and the pump photon is down-converted exactly back to the original two photons. Hence, if we choose the extended phase matching condition and the special condition, spectral entanglement between photons is avoided, and moreover, no population can be leaked to other spectral modes. So by tuning the Rabi oscillation such that $\frac{|\chi|NL}{\hbar}\sqrt{R}=\frac{\pi}{2}$, deterministic up-conversion can be achieved. 

The probability of successful conversion will be lowered if the extended phase matching condition and the special condition are not perfectly achieved. Lets suppose that some errors, $\epsilon_1$ and $\epsilon_2$, are allowed in the conditions, such that $k_{p}'=(1+\epsilon_1)\frac{k_{s}'+k_{i}'}{2}$ and $(1+\epsilon_2)^2L^{2}\gamma\sigma^{2}\left(k_{s}'-k_{p}'\right)\left(k_{p}'-k_{i}'\right)=\frac{1}{2}$. Assuming that $\frac{|\chi|NL}{\hbar}\sqrt{R}=\frac{\pi}{2}$, and having the set of reasonable parameters, $k'_s=5.6\times10^{-9} (s/m)$, $k'_i=5.2\times10^{-9} (s/m)$, and $\sigma=10^9 (Hz)$, then for $\epsilon_1=0.01$ and $\epsilon_2=0.01$, the probability of success is 0.9803; and for $\epsilon_1=0.001$ and $\epsilon_2=0.001$, the probability of success is 0.9998. Hence high conversion rates can be achieved if the errors in the extended phase matching condition and the special condition are reasonably small. 

\section{IV. Strong Chi-2 Non-Linearity Medium with Periodic Poling}
\begin{figure}[h]
\centerline{\psfig{figure=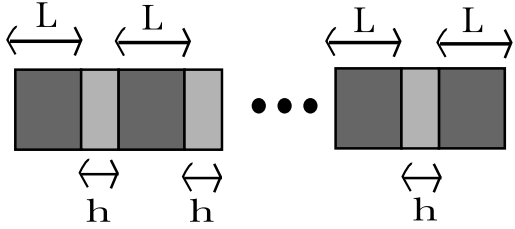,width=8.5cm}}
\caption{Schematic of $N$ slices of \X2 crystal of length $L$, with $N-1$ $\chi^{(1)}$ spacers of length $h$ in between.\\} \label{fig:Spacer}
\end{figure}

In the previous section, we analyzed the spectral effects of a $\chi^{(2)}$ medium obtained from many thin slices of crystal, such that the contributed non-linearity of each slice is small. This allows us to apply the simpler Taylor series to the unitary operator expansion, instead of the complicated Dyson series. However, guiding the modes through many thin slices, or many times through a single slice, could be experimentally challenging. In this section, we examine an alternative situation where we have a bulk medium comprising $N$ slices of $\chi^{(2)}$ crystal and with $N-1$ spacers in between. We prove that in the limit of sufficiently large $N$, it is valid to use the Taylor series to approximate the Dyson series. Figure~\ref{fig:Spacer} shows the schematic of $N$ slabs of $\chi^{(2)}$ crystal of length $L$ and $N-1$ slabs of $\chi^{(1)}$ spacers of length $h$ in between.

U'Ren et al~\cite{URen05} have shown that the interaction Hamiltonian of a medium consisting of $N$ \X2 crystals and $N-1$ periodically poled $\chi^{(1)}$ spacers has the expression:
\begin{align}
	\hat{\mathcal{H}}(t)=&\chi L\iiint_{-\infty}^{\infty} d\omega_{p}d\omega_{s}d\omega_{i}\hat{a}_{p}^{\dagger}(\omega_{p})\hat{a}_{s}(\omega_{s})\hat{a}_{i}(\omega_{i})e^{-i\Delta\omega t}\nn\\
	& \sinc\left(\frac{L\Delta k}{2}\right)e^{\frac{i L\Delta k}{2}}\frac{\sin(\frac{N\phi}{2})}{\sin(\frac{\phi}{2})}e^{\frac{i (N-1) \phi}{2}}+h.c.
\label{eq:Hs1}
\end{align}

\noindent where $\phi=L\Delta k+h\Delta\kappa$ and $\Delta\kappa=\kappa_p\nu_p-\kappa_s\nu_s-\kappa_i\nu_i$ is the phase mismatch introduced by each of the birefringent spacers, where $\kappa_j$ is the wavenumber for field $j$, taking into account the dispersion of the spacers. Rewriting the sin functions as sinc functions and approximating them as Gaussian functions gives:
\begin{align}
	\hat{\mathcal{H}}(t)=&\chi N L\iiint_{-\infty}^{\infty} d\omega_{p}d\omega_{s}d\omega_{i}\hat{a}_{p}^{\dagger}(\omega_{p})\hat{a}_{s}(\omega_{s})\hat{a}_{i}(\omega_{i})e^{-i\Delta\omega t}\nn\\
	& \sinc\left(\frac{L\Delta k}{2}\right)e^{\frac{i L\Delta k}{2}}e^{-\gamma\frac{N^2-1}{4}\phi^2} e^{i\frac{(N-1) }{2}\phi}+h.c.
\label{eq:Hs2}
\end{align}

If $h\kappa_p=L k_p'$, $h\kappa_s=L k_i'$ and $h\kappa_i=L k_s'$, then $\phi=L(2k_p'\nu_p-(k_s'+k_i')(\nu_s+\nu_i))$. Further applying the extended phase matching condition  gives $\phi=\eta\Delta\omega$, where $\eta=L(k_s'+k_i')$. Now the interaction Hamiltonian becomes:

\begin{align}
	\hat{\mathcal{H}}(t)=&\chi N L\iiint_{-\infty}^{\infty} d\omega_{p}d\omega_{s}d\omega_{i}\hat{a}_{p}^{\dagger}(\omega_{p})\hat{a}_{s}(\omega_{s})\hat{a}_{i}(\omega_{i})e^{-i\Delta\omega t}\nn\\
	& \sinc\left(\frac{L\Delta k}{2}\right)e^{\frac{i L\Delta k}{2}}e^{-\gamma\frac{N^2-1}{4}\eta^2\Delta\omega^2} e^{i\frac{(N-1)}{2}\eta\Delta\omega}+h.c.
\label{eq:Hs3}
\end{align}

Recall that the interaction Hamiltonian in equation~(\ref{eq:H1}) does not commute at different times, which has to do with the sinc function in the expression. In the limits $\eta\sqrt{N^2-1} \gg Lk'_j$, where $j$ is for all three of the modes, the sinc function in equation~(\ref{eq:Hs3}) is relatively flat in the domain where the Gaussian function is significant. This means that $\sinc\left(\frac{L\Delta k}{2}\right)\approx 1$ within the domain. Furthermore, the phase $\exp\left(\frac{i L\Delta k}{2}\right)$ can be neglected in this limit. Hence the interaction Hamiltonian tends to:

\begin{align}
	\hat{\mathcal{H}}(t)\approx&\chi N L\iiint_{-\infty}^{\infty} d\omega_{p}d\omega_{s}d\omega_{i}\hat{a}_{p}^{\dagger}(\omega_{p})\hat{a}_{s}(\omega_{s})\hat{a}_{i}(\omega_{i})e^{-i\Delta\omega t}\nn\\
	&e^{-\gamma\frac{N^2L^2}{4}(k'_s+k_i)^2\Delta\omega^2} e^{i\frac{NL}{2}(k'_s+k_i)\Delta\omega}+h.c.
\label{eq:Hs4}
\end{align}
  
\begin{figure}[h]
\centerline{\psfig{figure=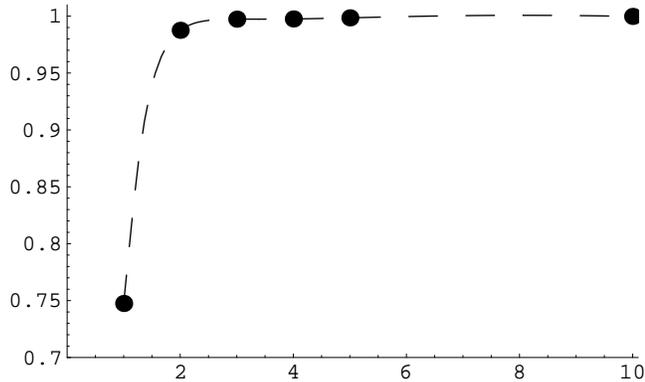,width=8.5cm}} 
\caption{Plot of $F_2$ against $N$\\} \label{fig:F2}
\end{figure}

Surprisingly, this interaction Hamiltonian commutes at different times and we may again use the Taylor series to calculate higher order terms of the unitary evolution. Using the extended phase matching condition, one of the three limits, $\sqrt{N^2-1}\eta \gg Lk'_p$, implies that  $N\gg\frac{\sqrt{5}}{2}\approx 1$, thus we expect that for $N\gg 1$ the Taylor series should give a good approximation to the Dyson series. To confirm this and determine how large $N$ has to be in practice such that the Taylor series gives a good approximation to the Dyson series, for various values of $N$, we calculate the parameter $F_2$ that we defined in equation~(\ref{eq:F2}). Figure~\ref{fig:F2} is the plot of $F_2$ against $N$. It shows that the fidelity between the 2nd order term of the Taylor series and the Dyson series is 0.998 for $N=5$ and continues to increase as $N$ gets larger. We have also checked that the phase difference between the two terms is negligible. Hence we argue that for a sufficiently large number of spacers placed between the \X2 crystals, we may use the Taylor series to approximate the Dyson series. Integrating the interaction Hamiltonian in equation~(\ref{eq:Hs3}) over time and simplifying the equation, we can write the Hamiltonian as:

\begin{align}
	\hat{\mathcal{H}}=&\chi N L\iiint_{-\infty}^{\infty} d\omega_{p}d\omega_{s}d\omega_{i}\hat{a}_{p}^{\dagger}(\omega_{p})\hat{a}_{s}(\omega_{s})\hat{a}_{i}(\omega_{i})\nn\\
	& \sinc\left(\frac{L\Delta k}{2}\right)e^{\frac{iL\Delta k}{2}}\delta(\Delta\omega) +h.c.\nn\\
	=&N\hat{H}
\end{align}

\noindent Since the Hamiltonian is for all slices of crystals and spacers, the state $|\psi_{odd}\ra=\sin(\hat{\mathcal{H}}/\hbar)|\psi_0\ra$. Hence, the profile of the up-converted photon is $|\psi_{odd}\ra=e^{i\theta}\frac{B}{\sqrt{R}}\sin\left(\frac{|\chi|NL}{\hbar}\sqrt{R}\right)|\psi_p\ra$ and the probability of up-conversion is $P(odd)=\frac{B^2}{R}\sin^2\left(\frac{|\chi|NL}{\hbar}\sqrt{R}\right)$, which are the same as the profile and probability that we have found in the case of many thin slices. So if we choose the extended phase matching condition and the special condition, which maintain spectrally separable photons and gives Rabi oscillation between $|\psi_p\ra$ and $|\psi_0\ra$ basis states, then by tuning the oscillation to $\frac{|\chi|NL}{\hbar}\sqrt{R}=\frac{\pi}{2}$, deterministic up-conversion can again be achieved.

\section{V. Bell Measurement and Quantum Gate schemes with strong Chi-2 non-linearity}

\begin{figure}[h]
\centerline{\psfig{figure=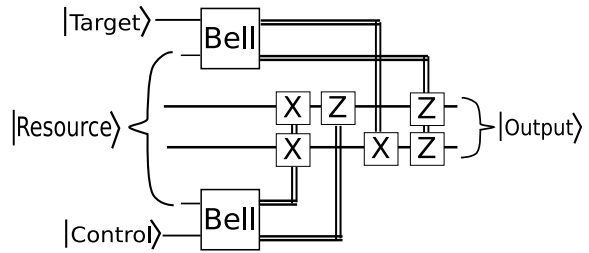,width=8cm}}
\caption{Schematic of the Gottesman and Chuang teleportation CNOT gate. The resource state $|Resource\rangle=((|HH\rangle+|VV\rangle)|HH\rangle+(|HV\rangle+|VH\rangle)|VV\rangle)/2$\\} 
\label{fig:cnot}
\end{figure}

Gottesman and Chuang~\cite{Gottesman99} showed that it is possible to build a CNOT gate by means of quantum teleportation and post-selection. Figure~\ref{fig:cnot} shows the CNOT gate scheme. The control qubit $|control\ra=a|H\ra+b|V\ra$ and the target qubit $|target\ra=c|H\ra+d|V\ra$ are two arbitrary single qubit states in polarization encoding. The resource state is the entangled state $((|HH\rangle+|VV\rangle)|HH\rangle+(|HV\rangle+|VH\rangle)|VV\rangle)/2$, which is prepared offline. The two input qubits are each sent to a separate device to perform a Bell measurement with the resource state. After measuring the input qubits and two of the four qubits of the resource state, the results are then used to perform single qubit operations on the two remaining output qubits. These single qubit operations can be done with optical waveplates. Gottesman and Chuang showed that this procedure teleports the input state to the output qubits with a CNOT operation applied between them.\\

\begin{figure}[h]
\centerline{\psfig{figure=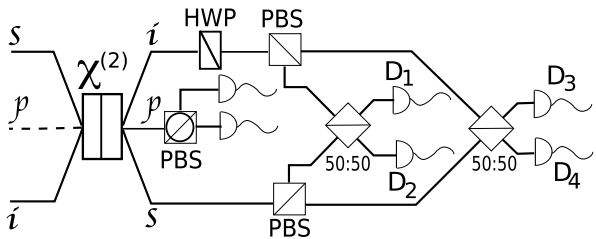,width=8cm}}
\caption{Schematic of the deterministic Bell measurement setup. The $\chi^{(2)}$ medium does TYPE II up conversion. The three modes $s, p, i$ are shown as different non-collinear spatial modes for the purpose of illustration. However, this is equivalent to collinear TYPE II conversion if one makes use of polarization beamsplitters and dichromatic beamsplitters to combine the three spatial modes into one before the conversion and then separate them again after the conversion.\\} \label{fig:Bell}
\end{figure}

The problem with implementing this gate with linear optics is that the Bell measurements only work $50\%$ of the time, giving a $\frac{1}{4}$ probability gate. Boosting the probability is possible but involves complex circuits~\cite{Knill01}. To perform the Bell measurement deterministically, we propose to use a sufficiently strong $\chi^{(2)}$ crystal along with linear optics as shown in Figure~\ref{fig:Bell}. The strong $\chi^{(2)}$ material consists of two slabs. One slab performs a Type II parametric up conversion for $|H\ra_s|V\ra_i$ and the other for $|V\ra_s|H\ra_i$. One may also construct a Bell measurement circuit using non-collinear Type I parametric up conversion in a similar fashion. The purpose of a Bell measurement is to distinguish the four different Bell states. To see how the scheme works, lets consider the cases of inputting the four Bell states. The two Bell states $(|H\ra_s|V\ra_i\pm|V\ra_s|H\ra_i)/\sqrt{2}$ are up-converted at the $\chi^{(2)}$ crystal to a pump photon with $(|H\ra_p\pm|V\ra_p)/\sqrt{2}$ states respectively by each slab. Thus measuring the polarization of the pump photon allows us to distinguish the two Bell states.
If the conversion fails, the two photons will pass through the crystal and the half wave plate (HWP) in the $i$ mode will turn the states into $(|H\ra_s|H\ra_i\pm|V\ra_s|V\ra_i)/\sqrt{2}$. These two states will have the photons bunched at either one of the two 50:50 beam splitters, giving two photons arriving at either one of the four detectors, $D_1$, $D_2$, $D_3$ or $D_4$, and thus signaling a failure event. Regarding the other two Bell states $(|H\ra_s|H\ra_i\pm|V\ra_s|V\ra_i)\sqrt{2}$, they simply pass through the $\chi^{(2)}$ crystal and the half wave plate in the $i$ mode will turn the state into $(|H\ra_s|V\ra_i\pm|V\ra_s|H\ra_i)\sqrt{2}$. The photons will then anti-bunch at the two 50:50 beam splitters. For $(|H\ra_s|V\ra_i+|V\ra_s|H\ra_i)\sqrt{2}$, the two photons will arrive at either $D_1$ and $D_3$ or at $D_2$ and $D_4$. For $(|H\ra_s|V\ra_i-|V\ra_s|H\ra_i)\sqrt{2}$, the two photons will arrive at either $D_1$ and $D_4$ or at $D_2$ and $D_3$. Thus, identifying the combination of flagged detectors allows us to distinguish the remaining two Bell states~\footnote{The second half of the Bell measurement circuit is exactly the non-deterministic Bell state analyzer that has been demonstrated with linear optics~\cite{Michler96}.}.

Although each Bell measurement uses two slabs of \X2 crystal, the probability of conversion is $P(odd)$ and not $P(odd)^2$ because there are only two photons undergoing one up-conversion process at any one time. Since two of the four Bell states are identified with a probability of $P(odd)$ and the other two are identified deterministically, the probability of a successful Bell measurement is therefore $\frac{1}{2}\left(1+\frac{B^2}{R}\sin^2\left(\frac{|\chi|NL}{\hbar}\sqrt{R}\right)\right)$. The CNOT gate comprises two Bell measurements that both need to succeed, so the probability of success of the gate is $P^2(odd)=\frac{1}{4}\left(1+\frac{B^2}{R}\sin^2\left(\frac{|\chi|NL}{\hbar}\sqrt{R}\right)\right)^2$. Given the extended phase matching condition and the special condition, and setting $\frac{|\chi| NL}{\hbar}\sqrt{R}=\frac{\pi}{2}$, the success rate of the Bell measurement and the gate becomes one. Hence by employing sufficiently strong $\chi^{(2)}$ crystals with the correct phase matching conditions, Bell measurements can be done with unit probability, and a deterministic CNOT gate can be achieved via teleportation. We note that Kim et al~\cite{Kim01} has proposed a similar Bell measurement scheme that requires two \X2 crystals and four slabs in total, to perform a Bell measurement. Since our scheme requires one crystal less, it gives a higher success probability over Kim et al's scheme if the conversion efficiency is less than one.

\section{IV. Conclusion}
In this paper, we have modeled the spectral effects of the dispersion of a \X2 crystal and pointed out that because the interaction Hamiltonian does not commute at different times, higher order terms of the unitary evolution of the state ought to be calculated using the Dyson series instead of the Taylor series. We have quantified the similarity between the second order term component of the state for the two series and estimated that the fidelity is $F_2=0.747$. Also, we have shown that the Dyson series induces spectral entanglement between the two input photons. We argued that this indicates that the conversion efficiency of a bulk \X2 crystal will be low. If the crystal is cut into many thin slices and well separated apart or has spacers between the slices, then the unitary operator can be calculated using the Taylor series. 
The two input photons can remain spectrally separable for the Taylor series, and thus high efficiency can be obtained. For the case of having $N-1$ $\chi
^{(1)}$ spacers periodically poled between $N$ slices of \X2 crystals, we argued that the interaction Hamiltonian approximately commutes at different times in the limit of large $N$ and demonstrated how the Taylor series is a good approximation to Dyson series in the limit by calculating $F_2$ for various values of $N$. We have estimated that with $N=5$, the fidelity $F_2$ is roughly 0.998 and the fidelity continues to improve as $N$ increases. We have derived the up-conversion rate to be $P(odd)=\frac{B^2}{R}\sin^2\left(\frac{|\chi|NL}{\hbar}\sqrt{R}\right)$ for both the many thin slices case and the spacers case. We found that if the extended phase matching condition $k_p'=\frac{k_s'+k_i'}{2}$ and the special condition $L^2\gamma\sigma^2(k_s'-k_p')(k_p'-k_i')=\frac{1}{2}$ are satisfied, and that $\frac{|\chi| NL}{\hbar}\sqrt{R}=\frac{\pi}{2}$, then the conversion rate becomes unity. We have further discussed how a Bell measurement can be done by employing a strong $\chi^{(2)}$ non-linearity to perform up-conversion on polarization encoded photons, and further implement it for constructing a teleportation-type CNOT gate. We calculated that the probability of success of the Bell measurement is $\frac{1}{2}\left(1+\frac{B^2}{R}\sin^2\left(\frac{|\chi|NL}{\hbar}\sqrt{R}\right)\right)$, and the success rate of the CNOT gate is $P^2(odd)=\frac{1}{4}\left(1+\frac{B^2}{R}\sin^2\left(\frac{|\chi|NL}{\hbar}\sqrt{R}\right)\right)^2$. Again, by having the extended phase matching condition and the special condition, and that $\frac{|\chi| NL}{\hbar}\sqrt{R}=\frac{\pi}{2}$, for both the many thin slices case and the spacers case, deterministic Bell measurement and CNOT gate can be constructed. In this paper, our discussion has focused on up-conversion, however a strong \X2 non-linearity has both up and down-conversion happening at the same time and thus the conditions and results that we derive here would also similarly apply to down-conversion.

\textbf{Acknowledgement}\\
We wish to thank Christine Silberhorn for helpful discussions. This work was supported by the Australian Research Council and the IARPA-funded U.S. Army Research Office Contract No. W911NF-05-0397.  We would also like to acknowledge support from the EU Integrated Project Qubit Applications (QAP) and the Ministry of Education, Culture, Sports, Science and Technology (MEXT) of Japan. P. Leung would also like to give special thanks to NII for their hospitality while a portion of this research was conducted there.

\end{document}